\documentstyle[epsfig]{camera}
\renewcommand{\baselinestretch}{0.9}
\begin{document}
\setcounter{topnumber}{5}
\def\topfraction{1.0}
\setcounter{bottomnumber}{5}
\def\bottomfraction{1.0}
\setcounter{totalnumber}{5}
\def\textfraction{0.0}
\def\floatpagefraction{1.0}
\def \met {{\,/\!\!\!\!E_{T}}}
\def \notbjet {{\check{j}}}
\def \bjet {b}
\def \deltaR {\Delta {\cal R}}
\newcommand{\et}{\rm E_T}
\def \code {\texttt}
\def \GeV {{\rm GeV}}
\def \metvec {{\rm\,/\!\!\!\!\vec{E}_{T}}}
\def \begineq {\begin{equation}}
\def \endeq {\end{equation}}
\def \scriptP {\mbox{${\cal P}$}}
\def \gothicP {\tilde{\scriptP}}
\def \ltapprox {\,\raisebox{-0.6ex}{$\stackrel{<}{\sim}$}\,}
\def \gtapprox {\,\raisebox{-0.6ex}{$\stackrel{>}{\sim}$}\,}
\def \Sherlock {{Sleuth}}
\def \hse {{\small{hse}}}
\newcommand {\abs}[1]{\mid \! #1 \! \mid}
\renewcommand{\thefootnote}{\arabic{footnote}}
\hyphenation{straight-forward}
\newcommand{\Pt}{$\rm P_T$}
\newcommand{\Pte}{$\rm P_T^e$}
\newcommand{\Et}{$\rm E_T$}
\newcommand{\Etj}{$\rm E_T^{j1}$}
\newcommand{\Etjj}{$\rm E_T^{j2}$}
\newcommand{\Etjjj}{$\rm E_T^{j2}+E_T^{j3}$}
\newcommand{\ppbar}{p{\bar p}}
\newcommand{\ttbar}{$t{\bar t}$}
\newcommand{\dytt}{Drell-Yan $\rightarrow \tau^+\tau^-$}
\newcommand{\dyll}{Drell-Yan $\rightarrow \ell^+\ell^-$}
\newcommand{\dyee}{Drell-Yan $\rightarrow e^+e^-$}
\newcommand{\ztt}{$Z^0 \rightarrow \tau^+\tau^-$}
\newcommand{\zee}{$Z^0 \rightarrow e^+e^-$}
\newcommand{\zmm}{$Z^0 \rightarrow \mu^+\mu^-$}
\newcommand{\DO}{$D\O$}
\newcommand{\pbarp}{p{\bar p}}
\newcommand{\etal}{{\em et al.}}
\newcommand{\mett}{\mbox{${\rm \not\! E}_{\rm T}$}}
\newcommand{\mettcal}{\mbox{${\rm \not\! E}_{\rm T}^{\rm cal}$}}

\title{SEARCHES FOR NEW PHYSICS AT TEVATRON}

\vspace{0.6cm}
\author{C. Pagliarone\footnote{{\it pagliarone@fnal.gov}\, or\, {\it carmine.pagliarone@pi.infn.it}}}
\organization{\vspace{-1.5cm} I.N.F.N. Pisa - {\it via
F.~Buonarroti, 2 - 56100 Pisa - ITALY\\
\vspace{1.3cm} (On behalf of CDF and D$\not$O Collaborations)}}
\maketitle

\vspace{-0.4cm}
\begin{abstract}
 This paper reviews the most recent results on searches for physics
beyond the Standard Model at Tevatron. Both the collider
experiments: CDF and D$\not$O are performing a large variety of
searches such as searches for scalar top and scalar bottom
particles, search for new gauge bosons, search for long-lived
massive particles and general searches for new particles decaying
into dijets. The results, summarized here, are a selection of what
obtained recently by both the collaborations using the Run II
data, collected so far.
\end{abstract}

\vspace{0.20cm}\section{Introduction}\

In the past two decades, Collider facilities have been places
where searches for physics beyond the Standard Model (SM), as well
as precise SM tests, have been pursued with great determination.
Since March 2001, CDF and D$\not$O~\cite{CDF}\cite{DO}, the two
Tevatron Experiments, are collecting data at the world highest
colliding beam machine energy ($\sqrt{s}=\,1.96\,\, TeV$). At the
present, April 2003, $\sim$ $150$ $pb^{-1}$ have been accumulated
by each experiment. Most of the analysis described in this paper
are based on an integrated luminosity ranging between $30$ and
$76$ $pb^{-1}$.

\vspace{0.20cm}\section{Search for new particles decaying to
dijets}\

 Many models predict the existence of new particles decaying, with
large branching fraction, in two partons. The dijet final states,
arising from such a decays, are often difficult to search because
of the large associated QCD background and the poor mass
resolution achievable. However, these disadvantages are partially
compensated by the large dijet statistics available. This allows
exclusion cross sections that are a small fraction of the total
dijet cross section. Figure~\ref{fig:dijet} (left) shows the
comparison between the Run I and Run II dijet spectrum. CDF-II has
obtained generic $95$\% C.L. upper limits on the cross sections
for narrow new resonances as a function of the mass. The obtained
upper limits have been compared to the production cross sections
for a variety of model as Axigluons, Flavor Universal Colorons,
Excited Quarks and $E_{6}$ diquarks. The extracted $95$\% C.L.
limits, obtained for the different models under consideration, are
quoted in Table~\ref{Tab:dijet}. Most of these limits extends
already Tevatron Run I exclusions into previously unexcluded mass
regions.

\begin{table}[t!]
\renewcommand{\baselinestretch}{1.40}
\centering
\begin{tabular}{c | l | l}  \hline\hline
{\bf MODEL}    & {\bf Run II} $\mathbf{(GeV/c^{2})}$ & {\bf Run I}
$\mathbf{(GeV/c^{2})}$
\\
\hline
{\bf Axigluon} & $ \mathbf{ 200 \, < \, M_{A\,} \,\, < 1130}$ &
 $\mathbf{ 200 \, < \, M_{A\,} \, < \,\,980}$\\
{\bf Excited Quarks} &
$\mathbf{200\,<\,M_{*\,}\,\,\,\,<\,\,\,760}$ & $\mathbf{200\,
<\,M_{*\,}\,\,\,<\,\,570}$\\ & & $\mathbf{580\,
<\,M_{*\,}\,\,\,<\,\,760}$\\
{\bf Color Octet Technirhos} & $ \mathbf{ 260 \, < \, M_{\rho\,}
\,\,\,\, <\,\,\, 640}$ &  $ \mathbf{ 260 \, < \, M_{\rho\,\,\,}
\,\, <\, 480}$
\\
$\mathbf{E_{6}}$ {\bf diquarks} &
$\mathbf{280\,<\,M_{E_{6}}\,<\,\,\,420}$ &
$\mathbf{290\,<\,M_{E_{6}\,}<\,420}$ \\
$\mathbf{W^{'}}$ & $\mathbf{300\,<\,M_{W^{'}}<\,\,\,410}$ &
$\mathbf{300\,<\,M_{W^{'}}<420}$ \\
{\bf Randall-Sundrum Gravitons} & $\mathbf{K/M_{Pl}}\,=\,0.3$,
$220 < M < 840$ & $\;\;\;\;\;\;\;\;\,\,\,\,\,\,\,\,\,$ $--$\\
\hline\hline
\end{tabular}
\renewcommand{\baselinestretch}{1.0}
\caption{Results on searches for new particles looking for dijet
narrow resonances; all the limits are expressed at $95$\% C.L.
(CDF-II).} \label{Tab:dijet}
\end{table}

\vspace{0.20cm}\section{Search for long-lived charged massive
particles}\

\begin{figure} [t!] \centering
\begin{minipage}{0.5\linewidth}
  \centering\epsfig{file=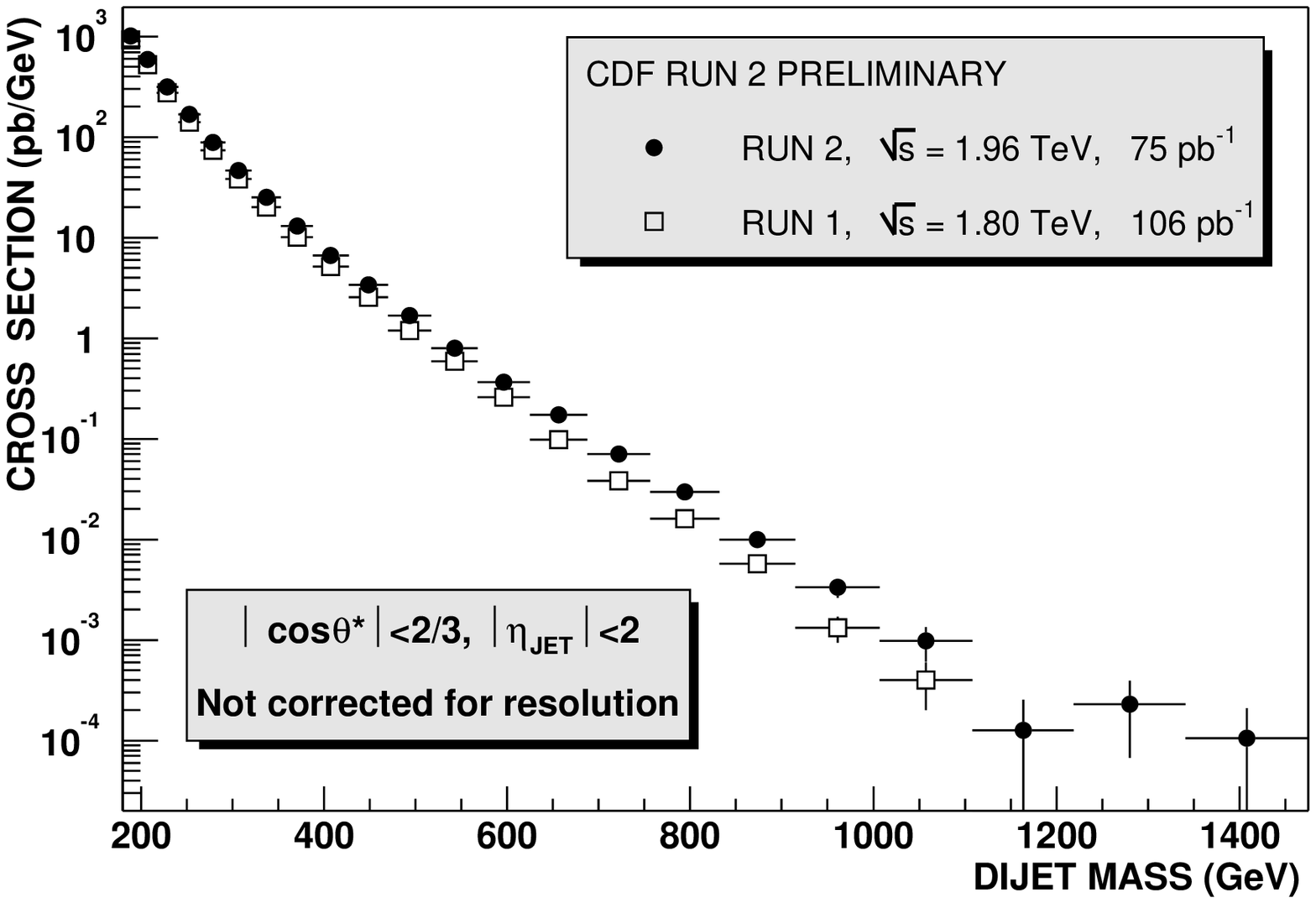,width=\linewidth}
\end{minipage}\hfill
\begin{minipage}{0.5\linewidth}
  \centering\epsfig{file=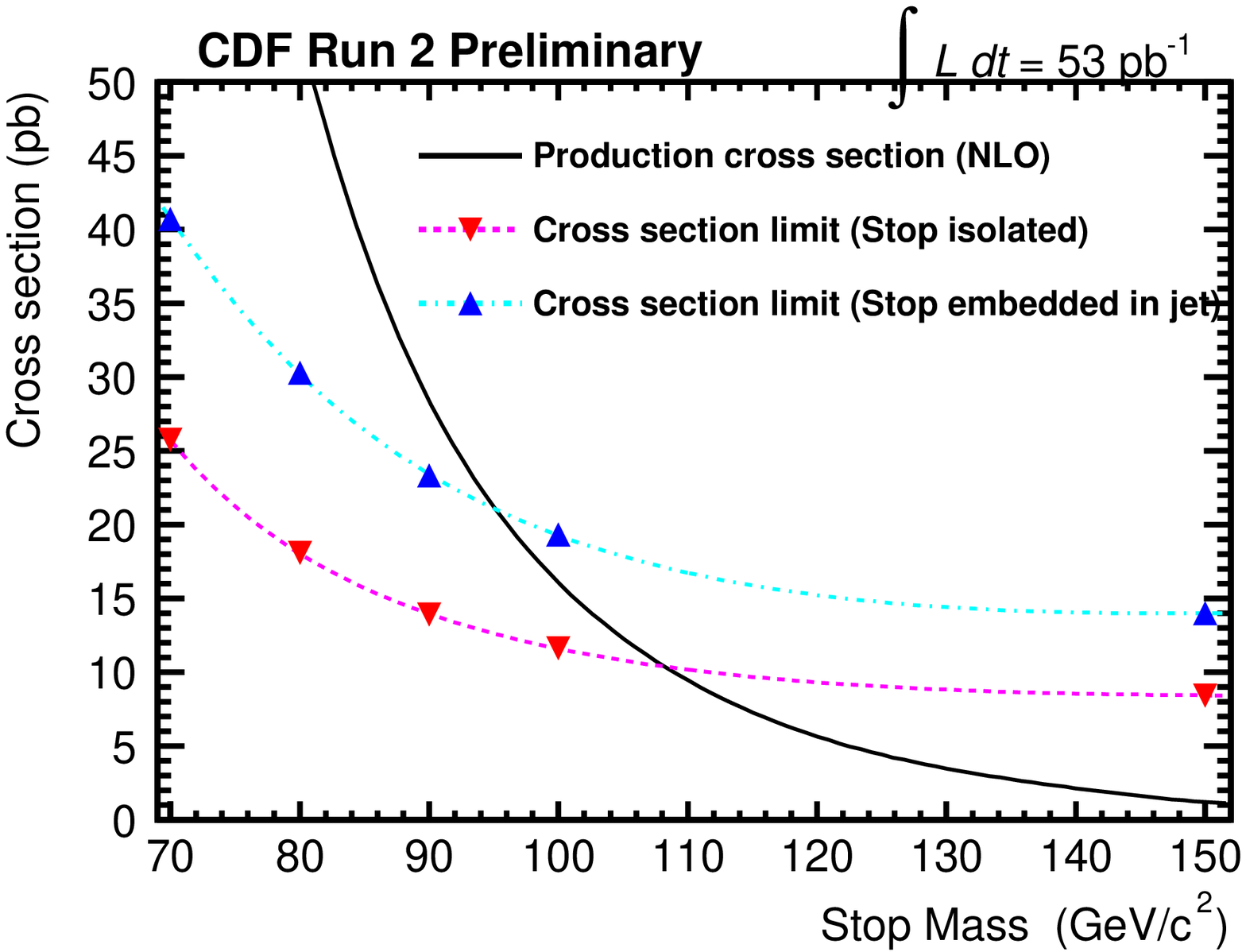,width=\linewidth}
\end{minipage}
\vspace{-.3cm} \caption{(Left) Comparison of Run I and Run II
dijet spectrum (CDF-II); (Right) NLO top squark production cross
section and the $95$\% C.L. limit as obtained from the search. The
resulting mass limit is $M(\tilde{t}_{1}) >\, 108$ $GeV/c^{2}$.
The NLO calculation assumes $\sqrt(s) =\, 1.96$ $TeV$, a
factorization scale $=\,M(\tilde{t}_{1})$ and the parton
distribution function CTEQ4M~\cite{CHAMPS3}. Varying the
factorization scale by $\pm1\,\sigma$ we get a variation of about
$\pm\,3$ $GeV/c^{2}$ in the obtained limit (CDF-II).}
  \label{fig:dijet}
\end{figure}

The existence of long-lived CHArged Massive Particles (CHAMPs) is
predicted in several theoretical models. Typically, CHAMPs are
expected to decay promptly. Most searches, consequently, attempt
to isolate distinctive decay signatures involving leptons, jets
and missing transverse energy ($\met$). There are, anyhow,
circumstances in which, one or more of these new massive
particles, can acquire a lifetime, that is long compared to the
typical time required to pass through an experimental setup such
as a high energy physics detector. If CHAMPs are long-lived enough
to escape from the detector, then, the signature will be
essentially a slow heavy charged particle, that can be detected
using the large $dE/dx$ or the large time of flight. Between the
models that predict the existence of long-lived CHAMPs, there are
SUSY extensions of SM with one compactified extra
dimension~\cite{CHAMPS1}. These models provide a highly specific
prediction. CHAMPs are a stop squarks with a mass around $200$
$GeV/c^{2}$. CHAMPs are also expected within the context of Gauge
Mediated SUSY Breaking models (GMSB)~\cite{CHAMPS2}. In this case,
if the scale of the SUSY breaking sector is sufficiently large,
the next to lightest SUSY particle (NLSP) can be stable. In
particular, third generation SUSY partners, as stau and stop, are
possible candidates for NLSP CHAMPs. CHAMPs can also arise in
models with $4$ generations in the case in which the fourth
generation is weakly coupled to the other three families. In this
case, the fourth generation quark is a CHAMP.\\ An analysis
searching for such long-lived CHAMPs has been performed by the
CDF-II collaboration. This search used the new CDF-II Time Of
Flight detector (TOF) that provides a better sensitivity to $\beta
\gamma$ compared to any deductible $dE/dx$ information. The
analysis used $75$ $pb^{-1}$ of data, collected with the
high-$P_{T}$ inclusive central muon trigger, selecting then
offline tracks with $P_{T}\,>\,40$ $GeV/c$. The information about
the time, in which the interaction occurred ($T_{0}$), is obtained
using low momentum tracks ($P_{T}\,<\,20\,\, GeV/c$). Then CHAMPs
are searched looking for tracks with large traversing time:
$\Delta T\,=\,T_{tracks} - T_{0}\,>\,2.5\,\,ns$. There are $7$
events that passed the analysis cuts. The number of expected
background events is $2.9$ $\pm$ $0.7$ $(Stat)$ $\pm$ $3.1$
$(Syst)$ and it has been evaluated using tracks with
$20<\,P_{T}\,<40$ $GeV/c$. The resulting mass limit, for the stop
scenario, is given in Figure~\ref{fig:dijet} (right) and
corresponds to a value of $M(\tilde{t}_{1})>\, 108\,\, GeV/c^{2}$.

\section{Search for Leptoquarks}\

Leptoquarks (LQ) are predicted in many extensions of the SM such
as Grand Unified Theories (GUT), Technicolor, {\it etc}. At
Tevatron, they can be pair produced through strong interactions:
$p \bar{p} \rightarrow \bar{LQ}LQ + X$ and decay in one of the
following final states: $\ell^{\pm}\ell^{\mp}q\bar{q}$ and
$\ell^{\pm} \nu q \bar{q}$ and $\nu \bar{\nu} q \bar{q}$. Both
Tevatron experiments searched in the past for $LQ$ by looking at
final states containing one or two leptons.

\subsection{First Generation Leptoquarks: dilepton-dijet channel}\

Here we report the latest CDF-II analysis performed by looking at
the channel $\bar{LQ}_{1} LQ_{1} \, \rightarrow \, eeqq$. Events
were selected by requiring the presence of two electrons with
$E^{e}_{T}\, >\, 25\,\,GeV$ and two jets with  $E^{jet}_{T}(1)\,
>\, 30\,\,GeV$ and $E^{jet}_{T}(2)\, >\, 15\,\,GeV$. Topological
cuts have been applied in order to reduce the SM background. No
events have been found using $72$ $pb^{-1}$ of data. The number of
background events expected from SM processes is $3.4$ $\pm$ $3.0$.
The corresponding $95$\% C.L. limit is : $M(LQ_{1})\,> 230\,\
GeV/c^{2}$ (see Figure~\ref{fig:LQ} left).\

Also D$\not$O searched for First Generation Leptoquarks by
studying the same decay channel and following a similar approach.
The number of events observed, in the analysis, were $0$ with an
expected background of $0.08\,\pm\,0.02$. The obtained limit at
$95$\% C.L. is $M(LQ_{1})\,> 179\,\, GeV/c^{2}$ and already
improve the previous Run I D$\not$O result.

\subsection{First Generation Leptoquarks: jet+ $\met$ channel}\

\begin{figure} [t!] \centering
\begin{minipage}{0.5\linewidth}
  \centering\epsfig{file=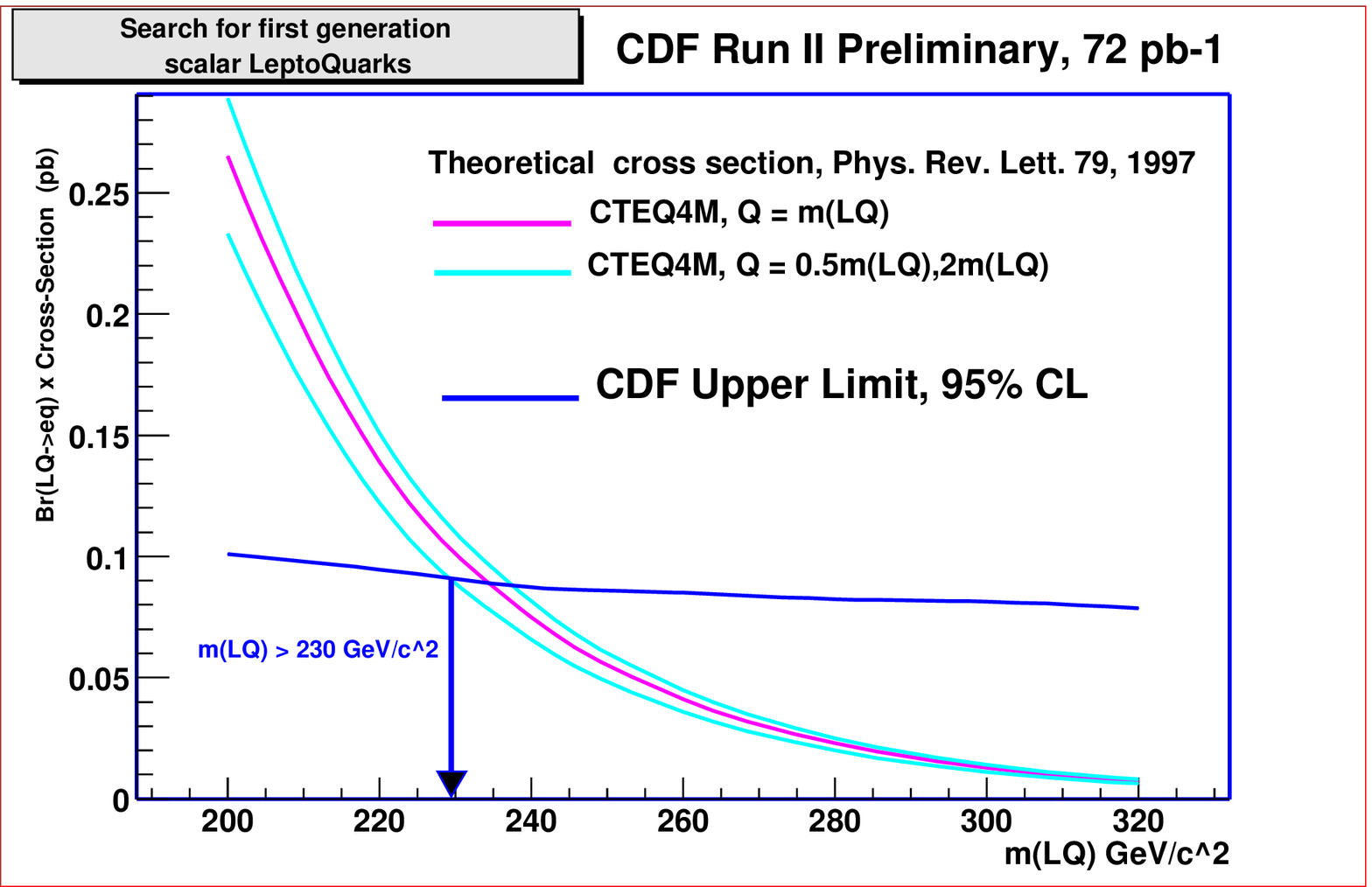,width=\linewidth}
\end{minipage}\hfill
\begin{minipage}{0.5\linewidth}
  \centering\epsfig{file=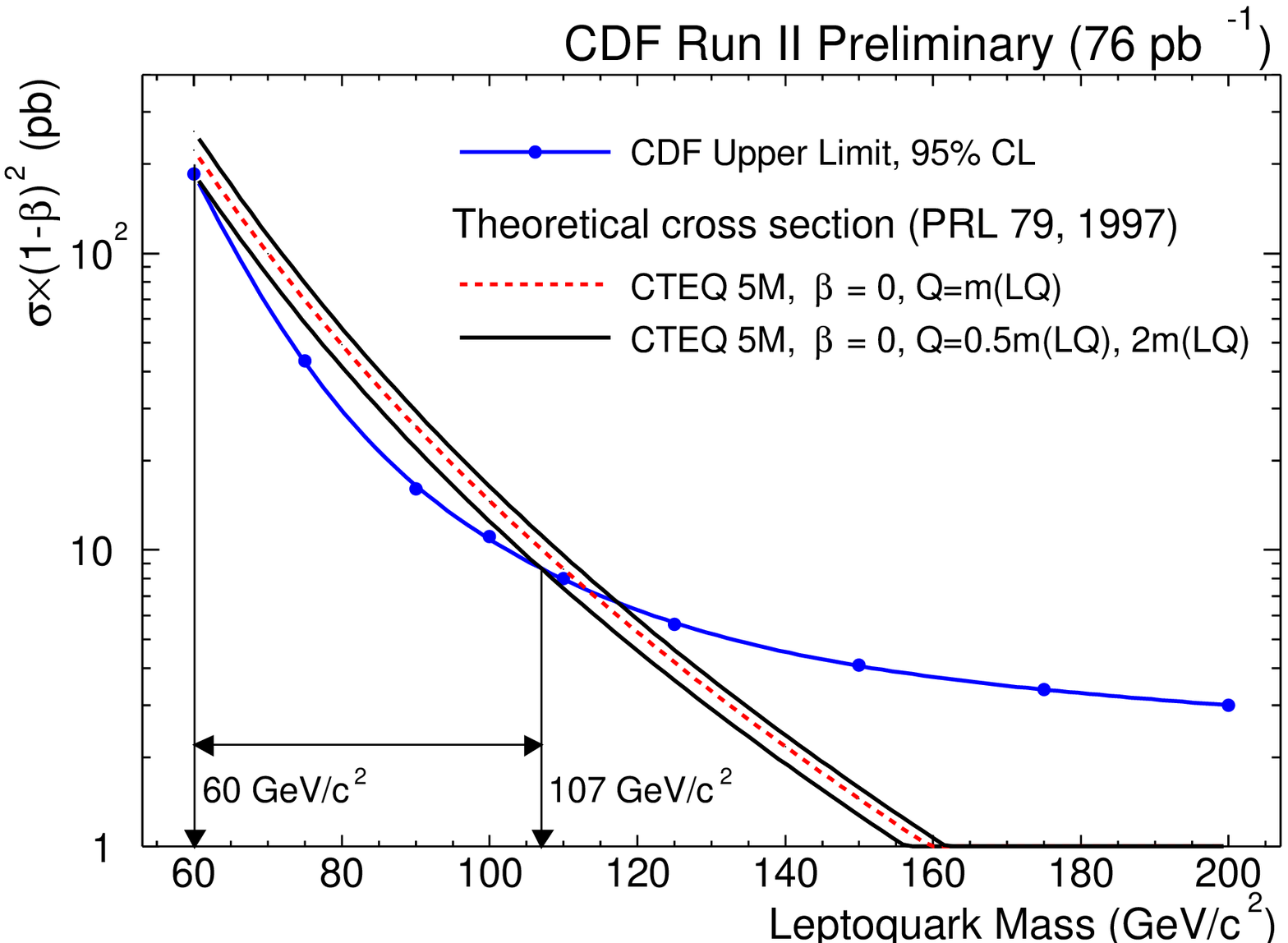,width=\linewidth}
\end{minipage}
\vspace{-.3cm}\caption{ (Left) CDF-II limit on the cross section
as a function of $M(LQ_{1})$ compared with the theoretical
expectations calculated at NLO accuracy. At the intersection point
between experimental and theoretical curves we find a lower limit
on $M(LQ_{1})$ at $166\,\,GeV/c^{2}$. ($\beta\,=\,\,0.5$); (Right)
Search for $p \bar{p} \rightarrow LQ_{1} \bar{LQ}_{1}$, with
$LQ_{1} \rightarrow q + \nu$. The signature for this process is
large missing transverse energy and two energetic jets (CDF-II).}
  \label{fig:LQ}
\end{figure}

CDF-II searched also for First Generation Leptoquarks production
considering the $LQ_{1}$ decaying into neutrino and quark ($LQ_{1}
\rightarrow  \nu \bar{\nu} q \bar{q}$) which yields missing
transverse energy and several high-$E_{T}$ jets in the final
state. The analysis have been performed using $76$ $pb^{-1}$ of
data. The number of events observed in the signal region is $42$
with an expected background of $42$ $\pm$ $11$. Therefore no
evidence for leptoquark production has been observed. The
resulting $95$\% C.L. limit on the cross section times squared
branching ratio is given in Figure~\ref{fig:LQ} (right). We
exclude the mass interval:
$60\,GeV/c^{2}\,<\,M(LQ_{1})\,<107\,GeV/c^{2}$

\subsection{Second Generation Leptoquarks}\

Another analysis performed by D$\not$O collaboration, using $30$
$pb^{-1}$ of data, is the search for Second Generation Leptoquarks
decaying trough the channel: $LQ_{2} \bar{LQ}_{2} \rightarrow
\mu^{+} \mu^{-} q q$. Events were selected by requiring the
presence of two opposite sign muons with $P^{\mu}_{T}\,>\,\,
15\,\,GeV/c$ and two jets with $E^{jet}_{T}\, >\, 20\,\,GeV$. The
dominant background for this process is $ Z/\gamma^{*} \rightarrow
\mu^{+} \mu^{-}$ associated with two jets. The number of events
that passed the cuts is $0$; the corresponding $95$\% C.L. limit
on $LQ_{2}$ is $M(LQ_{2})<\,157\,\, GeV/c^{2}$. The previous Run I
limit, based on the full statistics, was:  $M(LQ_{2})<\,200\,\,
GeV/c^{2}$.

\vspace{0.20cm}\section{Search for doubly-charged Higgs
(dielectron channel)}\

Doubly-charged Higgs particles ($H^{\pm\pm}$) are expected in
several theoretical frameworks. Such particles are members of
Higgs triplets that occur in theories with extensions to the Higgs
sector of the SM~\cite{HH1}, left-right (LR) symmetric
models~\cite{HH2} and SUSY LR symmetric models~\cite{HH3}. Because
of the charge conservation, the $H^{\pm\pm}$ must decay to either
two same-sign leptons, two same-sign W bosons, two same-sign
singly-charged Higgs bosons or a singly-charged Higgs and a W
boson. In particular, SUSY LR models predict low-mass $H^{\pm\pm}$
($\sim 100$ $GeV/c^{2}$ to $\sim 1$ $TeV/c^{2}$). CDF-II have
searched for these particles decaying into two leptons using $91$
pb$^{-1}$ of data, collected between March 2002 and January 2003.
Even if SUSY LR models provides motivations to search for low mass
doubly-charged Higgs particles, a search in the same sign (SS)
di-electron data is sensitive to any doubly charged particle
decaying to electrons. CDF-II searched for same-sign good central
electrons passing the following kinematical cuts:
$E^{e}_{T}\,>\,30$ $GeV$ and track $P_{T}\,>\,10$ $GeV/c$. The
mass resolution is about 3\% of $M(H^{\pm\pm})$. The search region
is defined as a mass window of $\pm\,10\%$ . The search has been
performed using a SS sample in the mass region above the $Z$ pole
($M_{e^{\pm}e^{\pm}}>\,100$ $GeV/c^{2}$) to avoid possible
contamination coming from Z events with a charged leg
misinterpreted. No events have been observed in the search region
of same sign mass above $100$ $GeV/c^{2}$. These results provide a
$95$\% C.L. limit for pair-production of doubly-charged particles.

\vspace{0.20cm}\section{Search for New Gauge Bosons}\

\begin{figure} [t!] \centering
\begin{minipage}{0.5\linewidth}
  \centering\epsfig{file=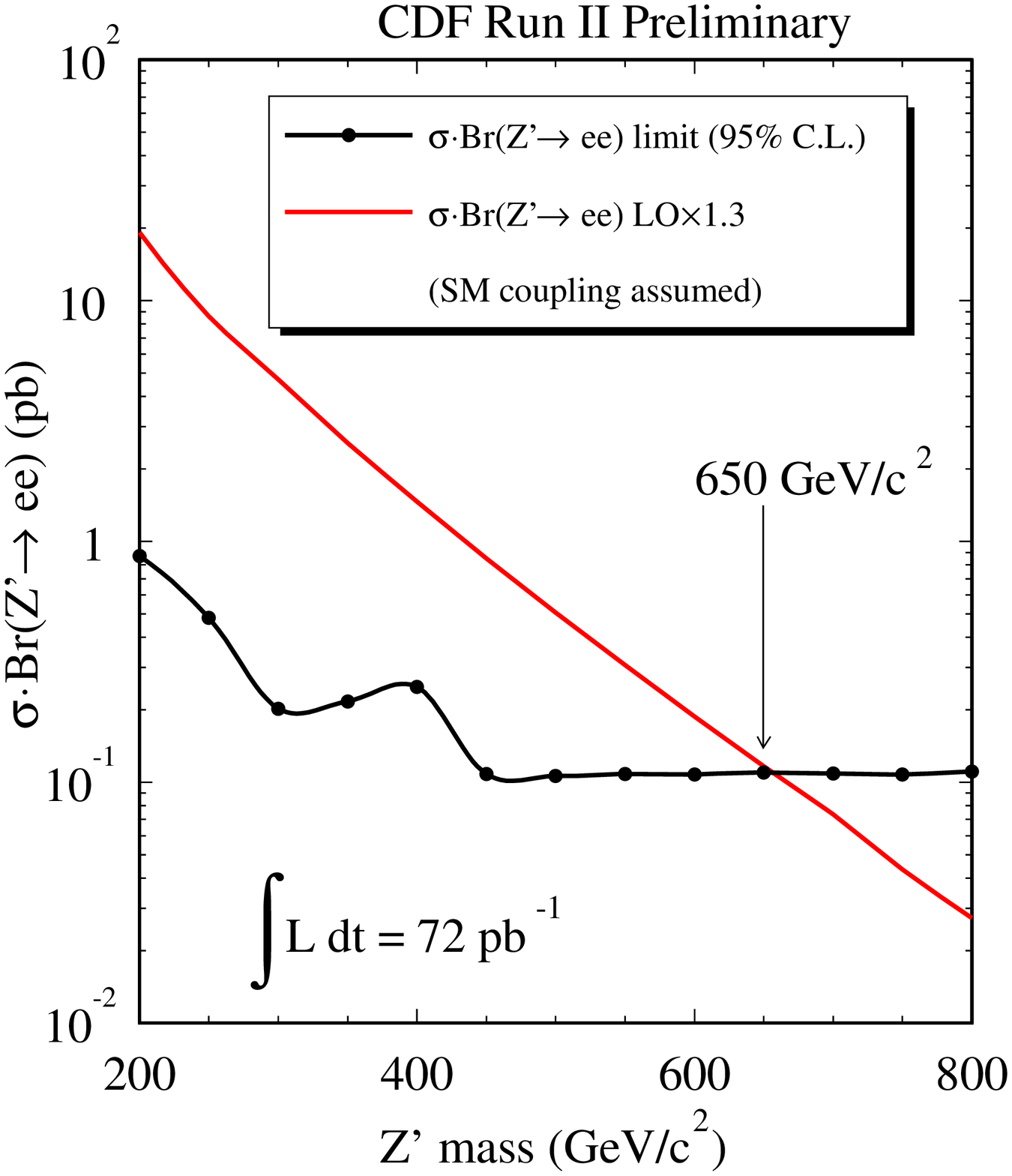,width=\linewidth}
\end{minipage}\hfill
\begin{minipage}{0.5\linewidth}
  \centering\epsfig{file=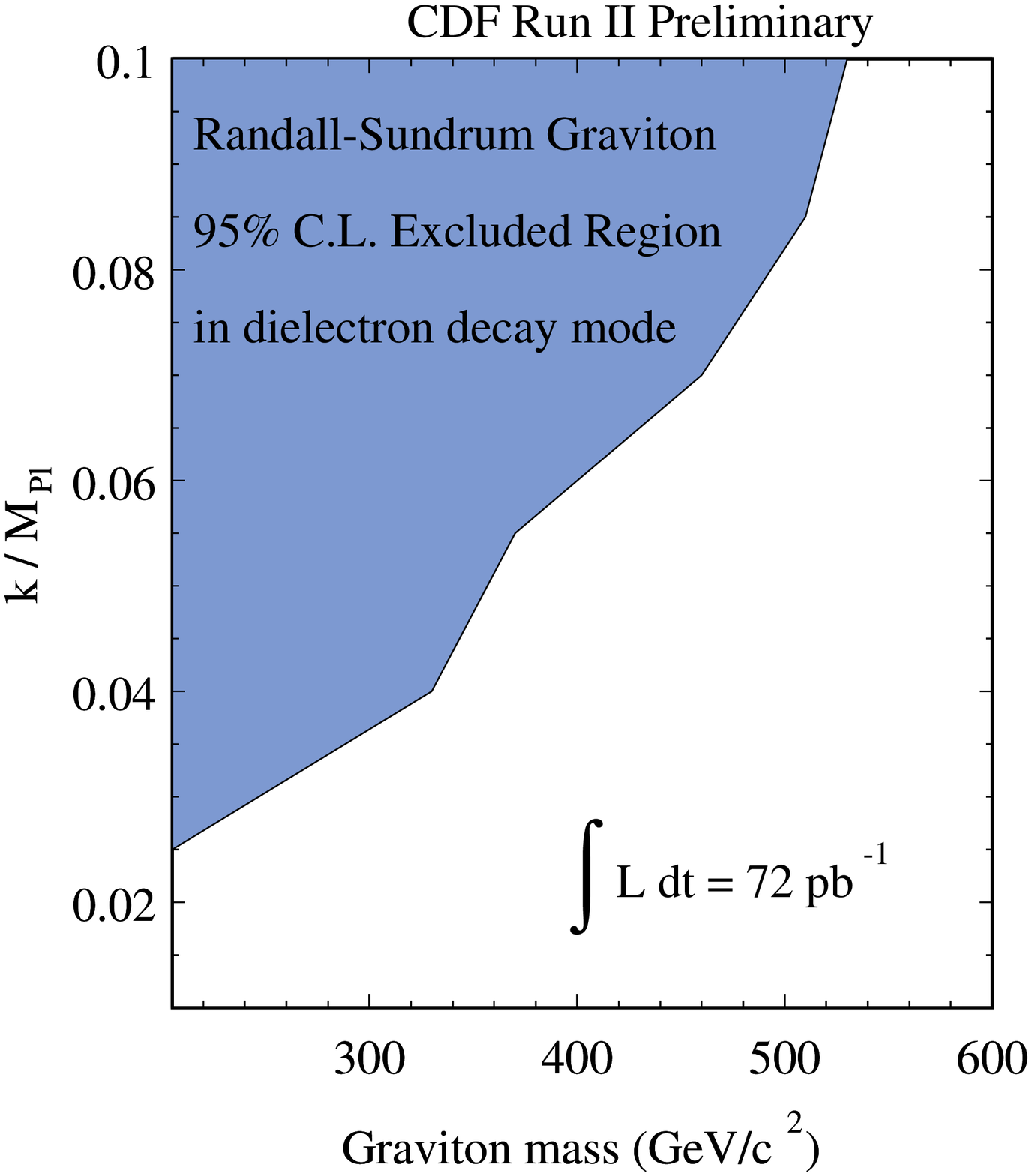,width=\linewidth}
\end{minipage}
\vspace{-.3cm}
  \caption{(Left) CDF-II search for new gauge bosons: $95$\% C.L. limit
  on $Z^{'} \rightarrow e^{+}e^{-}$ search; (Right)
  Randall-Sundrum Graviton excluded region.}
  \label{fig:Z_RS}
\end{figure}

CDF and D$\not$O searched for new neutral gauge boson $Z^{'}$ and
Randall-Sundrum gravitons decaying into dileptons. From an
experimental point of view such particles are generally high mass
states produced by $q \bar{q}$ annihilation and decaying into a
pair of opposite sign leptons. The primary observable effect will
be an anomalous dilepton production at large invariant masses
enhancing the Drell-Yan cross section. From a theoretical point of
view, the existence of neutral gauge bosons in addition to the
Standard Model Z is predicted in many extension of the SM. These
models specify the strengths of the couplings of the $Z^{'}$ to
quarks and to leptons. In the Randall-Sundrum model, Kaluza-Klein
(KK) excitations can be produced as resonances enhancing the SM
Drell-Yan production cross section at large mass. The CDF analysis
is based on $72$ $pb^{-1}$ of data collected between March 2002
and January 2003 using high-$P_{T}$ electron data. Events where
selected by requiring two electrons with :
$E^{e}_{T}\,>\,25\,GeV/c$. Specific cuts are applied in order to
reduce the QCD $W\,+\,jets$ background. The corresponding CDF-II
limit on the $ Z^{'}$ mass is $650$ $GeV/c^{2}$ as shown in
Figure~\ref{fig:Z_RS} (left). The D$\not$O limit on the same
channel is $455$ $GeV/c^{2}$. The limit obtained for the
Randall-Sundrum scenario is given in Figure~\ref{fig:Z_RS}
(right).

\vspace{0.20cm}\section{Physics with Taus in the final state}\

The study of processes containing $\tau$ leptons in the final
state will play an important role at Tevatron Run II. Such final
states is relevant both for electroweak studies and measurements
as well as in searches for physics beyond the Standard Model. In
particular to search for a variety of new physics scenarios such
as SUSY, SUSY with $\mathcal{R_{P}}$-parity violation (RPV), SUSY
with Bilinear parity violation (BRPV) or models with the violation
of lepton flavor (LFV). For this purpose a new set of triggers
have been implemented able to select events containing tau
candidates in the final state.

\vspace{0.20cm}\subsection{The CDF-II Tau Triggers}\

The CDF-II $\tau$ Triggers are a set of Triggers integrated into
all 3 levels of the general CDF-II Trigger system~\cite{CDF}. At
the present CDF-II has five different $\tau$ triggers operating:
\begin{itemize}
\item Central Muon Plus Track;
\item CMX Muon Plus Track;
\item Central Electron Plus Track;
\item Di-Tau Trigger;
\item Tau + $\not\!\!\!E_{\rm T}$;
\end{itemize}
These Triggers were installed in the CDF-II trigger tables in
January 2002. Naturally, the design of these triggers has evolved
in time. At the present they are all working properly collecting
data in stable, non prescaled way. In particular the lepton plus
track triggers are a class of low momentum dilepton triggers able
to select events containing charged leptons, including $\tau$'s,
in the final state.

\noindent As taus in $\sim\,35$ \% of cases promptly decay into
leptons and the rest of times in hadrons, then dilepton events,
where both leptons are $\tau$'s, can be identified by accessing
both purely leptonic di-$\tau$ decays: $\,\tau_{e} \,\tau_{e}\,$,
$\,\tau_{e} \, \tau_{\mu}\,$ or mixed leptonic-hadronic di-$\tau$
decays: $\,\tau_{e} \,\tau_{h}\,$ or $\,\tau_{\mu} \,\tau_{h}\,$.
Then the full accessible final states are: $e\,e$, $\,e \,\mu$,
$\,e \,\tau_{h}$, $\,\mu \,\mu $, $\,\mu \,\tau_{h}$.
Hadronic decays of taus result in jets that must be distinguished
from jets arising from QCD processes. In this case the
``$\tau$-jetiness'' is ensured by the isolation criteria applied
around the second track at Level 3. As a corollary, this prevents
the track from being a product of a light quark or heavy flavored
quark jet. At the present many analysis are underway using data
collected with these triggers. Results will be released soon.

\vspace{0.20cm}\section{Conclusions}\

The search for new physics is a primary goal for a hadron
collider. CDF and D$\not$O following an important Run I tradition
are pursuing with great determination a large variety of exotic
searches. Many Run I limits have been already enhanced using the
data collected so far.

\vspace{0.20cm}\section{Acknowledgements}\

I would like to thank the organizers of\, IFAE\, 2003 conference
for the stimulating atmosphere and well organized program of
talks. This paper is dedicated to memory of my grandmother.



\end{document}